cd

\documentclass[12pt]{iopart}
\usepackage [dvips]{graphics, color}
\usepackage{iopams,wrapfig}
\usepackage{xspace}
\usepackage{subfig}

\newcommand{\sNNtwohundred}{$\sqrt{s_{NN}}$ = $200$ $GeV$\xspace}
\newcommand{\sNNsixtytwo}{$\sqrt{s_{NN}}$ = $62$ $GeV$\xspace}
\newcommand{\sNN}{$\sqrt{s_{NN}}$\xspace}
\newcommand{\stdassoc}{$1.5$ $GeV/c$ $<$ $p_T^{associated}$ $<$ $p_T^{trigger}$\xspace}
\newcommand{\stdtrig}{$3.0$ $<$ $p_T^{trigger}$ $<$ $6.0$ $GeV/c$\xspace}
\newcommand{\pttrig}{$p_T^{trigger}$\xspace}
\newcommand{\ptassoc}{$p_T^{associated}$\xspace}
\newcommand{\npart}{$N_{part}$\xspace}
\newcommand{\pT}{$p_T$\xspace}
\newcommand{\jet}{\textit{Jet}\xspace}
\newcommand{\ridge}{\textit{Ridge}\xspace}
\newcommand{\kaon}{$K^0_S$\xspace}
\newcommand{\lam}{$\Lambda$\xspace}

\newcommand{\lamalam}{$\Lambda+\bar{\Lambda}$\xspace}
\newcommand{\xibary}{$\Xi$\xspace}

\newcommand{\pp}{$p+p$\xspace}
\newcommand{\Cu}{$Cu+Cu$\xspace}
\newcommand{\Au}{$Au+Au$\xspace}
\newcommand{\dAu}{$d+Au$\xspace}

\newcommand{\MeV}{MeV/c\xspace}
\newcommand{\GeV}{GeV/c\xspace}
\newcommand{\dphi}{$\Delta\phi$\xspace}
\newcommand{\deta}{$\Delta\eta$\xspace}
\newcommand{\vtwo}{$v_2$\xspace}
\newcommand{\nearside}{near-side\xspace}

\begin{document}

\article[System and energy dependence of particle correlations]{Quark Matter}{System and energy dependence of strange and non-strange particle correlations in STAR at RHIC}
\author{C E Nattrass for the STAR Collaboration}
\address{WNSL, 272 Whitney Ave., Yale University, New Haven, CT 06520, USA}
\ead{christine.nattrass@yale.edu}


\begin{abstract}
Two-particle high-\pT triggered correlations in \Cu and \Au collisions at \sNNsixtytwo and \sNNtwohundred in STAR at RHIC are presented.  The \npart, \ptassoc, and \pttrig dependence of the yield per trigger in the \jet and \ridge components of the \nearside is investigated for h, \kaon, \lam, and \xibary trigger and associated particles.  The system and energy dependence of these components is a potentially powerful tool to distinguish between models used to describe their production.

\end{abstract}

\section{Data analysis and results}
Previous studies in \Au collisions at \sNNtwohundred demonstrated that the near-side peak in high-\pT triggered correlations can be decomposed into two structures.  The \jet is narrow in both azimuth (\dphi) and pseudorapidity (\deta), similar to what is observed in \dAu, while the \ridge is narrow in azimuth but broad in pseudorapidity.  The \jet component is similar to that expected from vacuum fragmentation, whereas the \ridge has properties similar to the bulk\cite{Jana,Joern}.  Comparing data from \Au and \Cu collisions at \sNNsixtytwo and \sNNtwohundred tests whether these conclusions are robust.

Several mechanisms have been proposed for the production of the \ridge\cite{LongFlow,MomentumKick,Reco,Sergei}.  These models have yielded few calculations which can be directly compared to data, in part because of the number of factors which must be considered when theoretically calculating the experimentally measured quantitites.  These data should provide a good test of models for the production of the \jet and \ridge because trends expected with changing collision energy and in nuclei collided in a given model should be easier to calculate theoretically.

The distribution of associated particles relative to a high-\pT trigger particle in azimuth (\dphi) and pseudorapidity (\deta) is normalized per trigger.  The elliptic flow (\vtwo) modulated background is subtracted assuming a two-component model and using the Zero-Yield-At-Minimum (ZYAM) method\cite{phenixZYAM}.  The background is set by averaging over three points near the minimum at \dphi $\approx$ 1.  The validity of this assumption is checked by comparing to the ZYAM method with one point and by allowing the background level to vary in a fit.  All correlations are corrected for the detector efficiency of the associated particle.  The \ridge yield is given for $|\Delta\eta|<\xspace 1.75$ and $|\Delta\phi|<\xspace 1$ and the \jet yield is extracted in the range $|\Delta\eta|<\xspace 0.75$.  The \kaon, \lam, and \xibary are identified by reconstruction of their decay vertices in the STAR TPC.  \cite{Me} provides a more complete description of both yield extraction and particle identification.  Unless otherwise noted, \stdassoc and \stdtrig.  The data from \Au collisions at \sNNtwohundred are from \cite{Jana}.

A systematic error due a detector effect, discussed in greater detail in \cite{Marek}, which causes lost tracks at small \dphi and small \deta has not yet been taken into account.  This effect, on the order of 10\%, is greater in higher multiplicity environments, at lower \ptassoc and \pttrig, and for particles identified by the reconstruction of decay vertices, and therefore reduces the \jet yield more for \kaon and \lam than h and more in \Au than in \Cu collisions.  The degree of correlation between systematic errors on the \ridge yield for different particle species due to \vtwo is still being studied.  These systematic errors affect both identified trigger and associated particle studies.


\begin{figure}
\begin{center}
\hspace{0.0cm} 
\rotatebox{0}{\resizebox{15.cm}{!}{
	\includegraphics{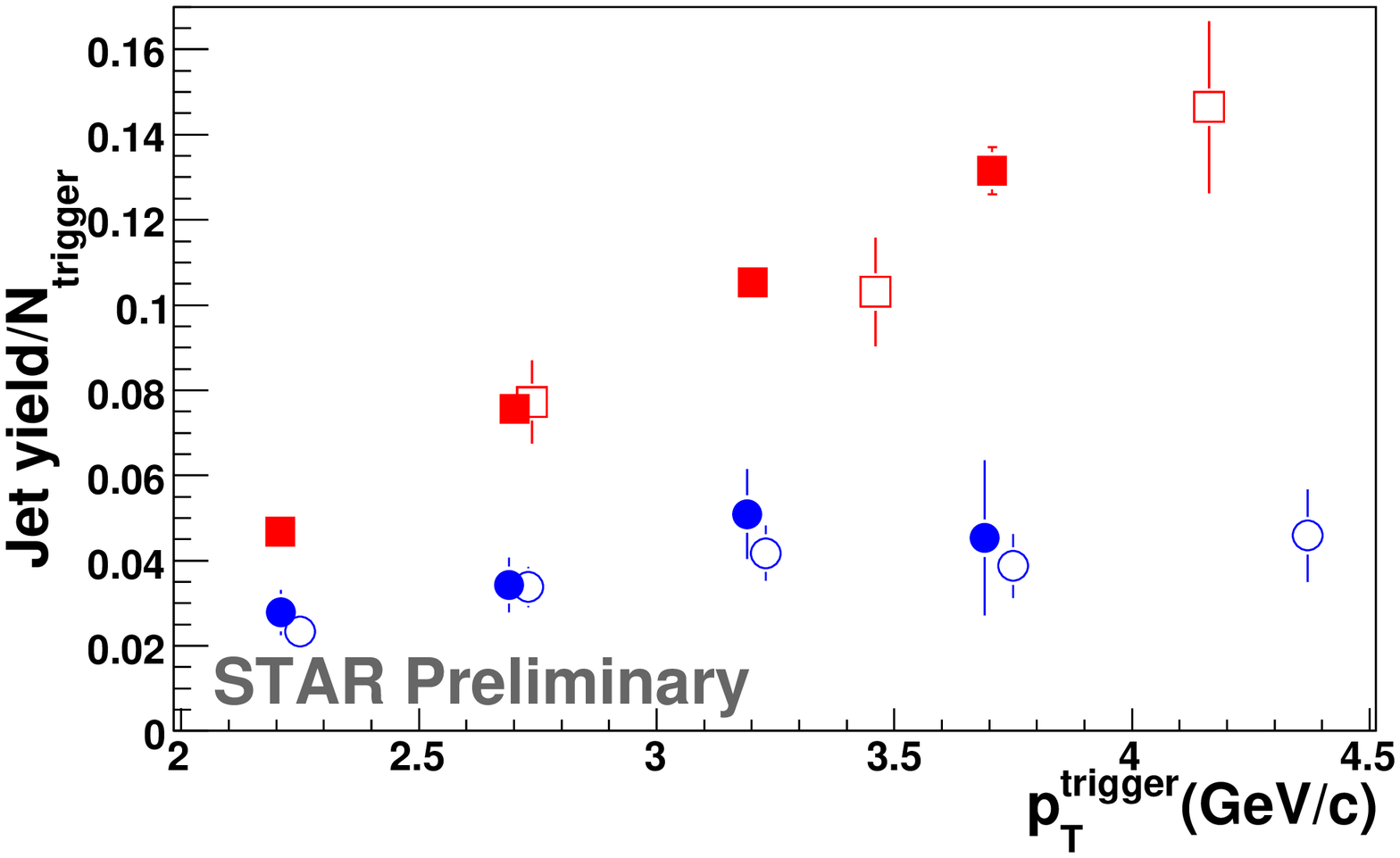}
	\includegraphics{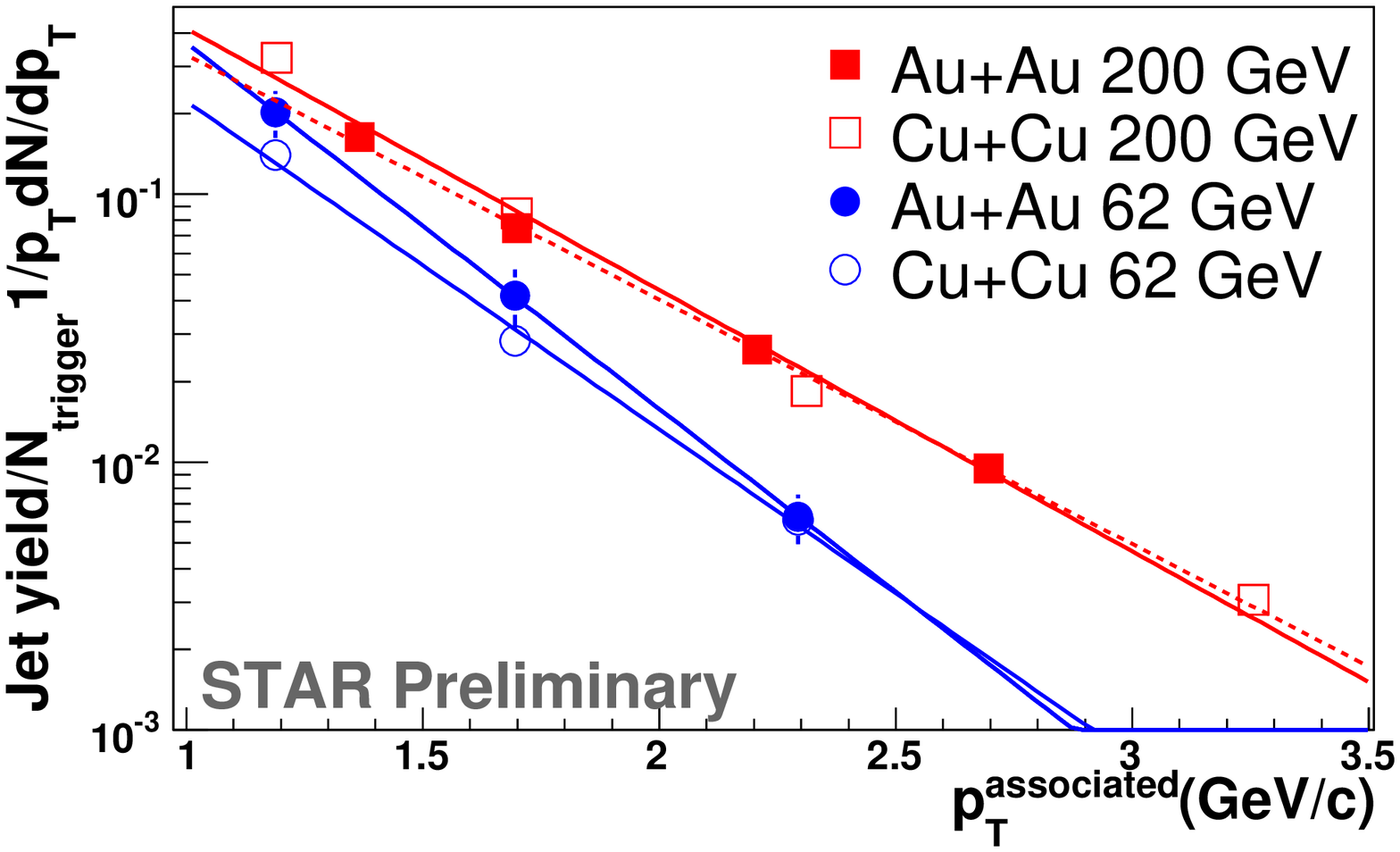}
}}
\textbf{}

\vspace{-.7cm} 
\begin{tabular*}{15.7cm}{ @{\extracolsep{6.5cm}} l l @{\extracolsep{\fill}}}
\hspace{0.5cm} \textbf{(a)} & \textbf{(b)}\\
\end{tabular*} 

\hspace{0.0cm} 
\rotatebox{0}{\resizebox{15.cm}{!}{
	\includegraphics{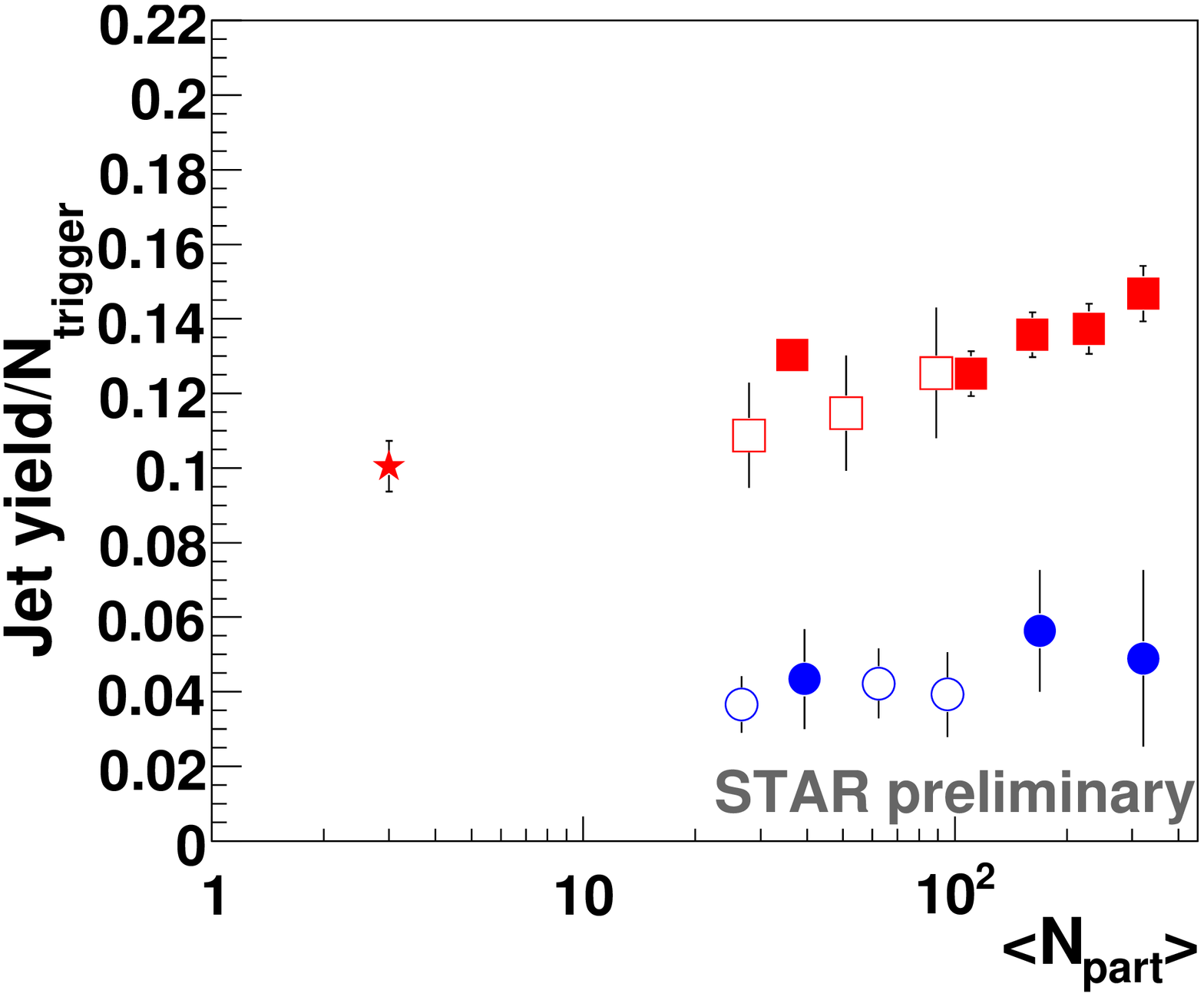}
	\includegraphics{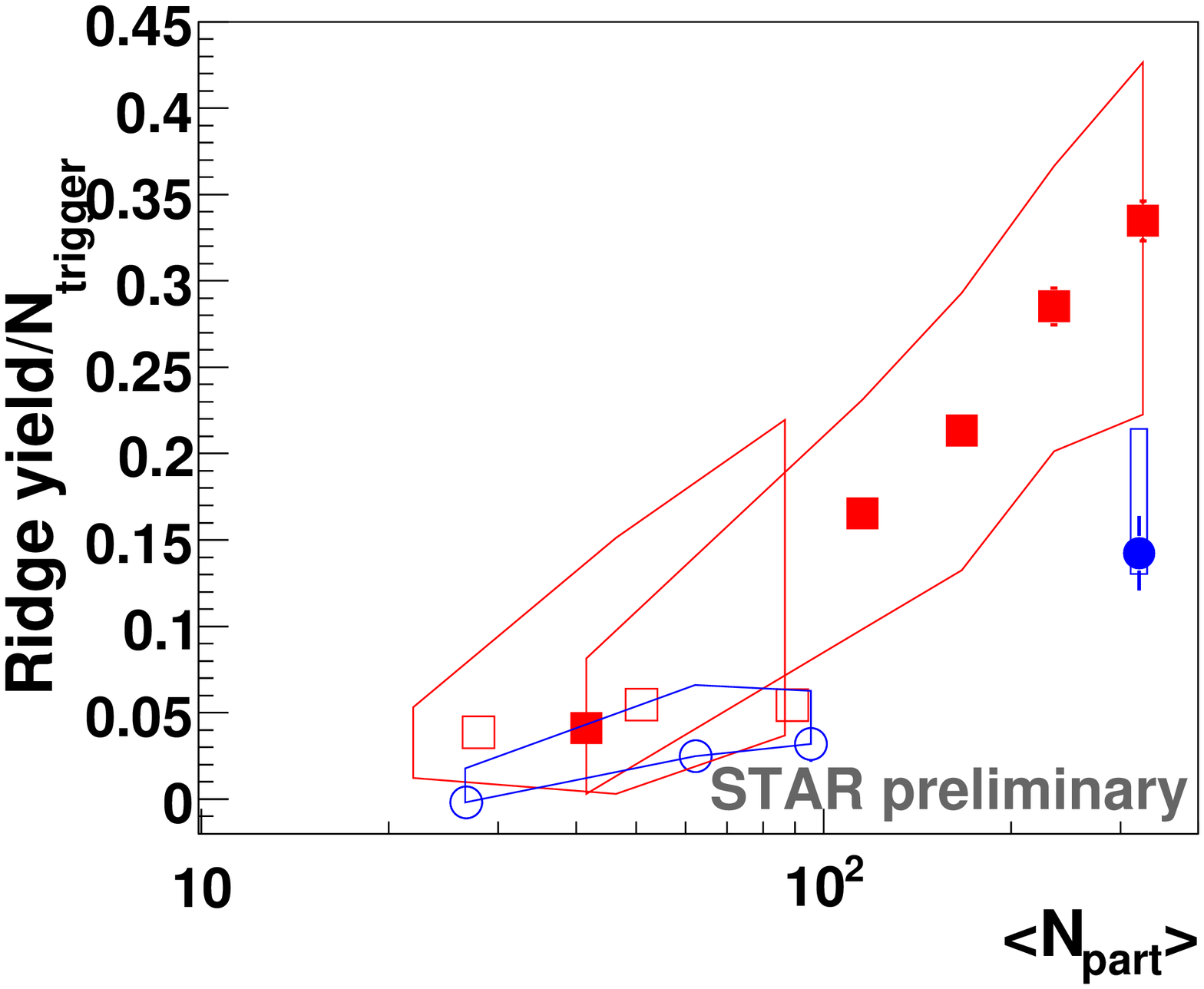}
	\includegraphics{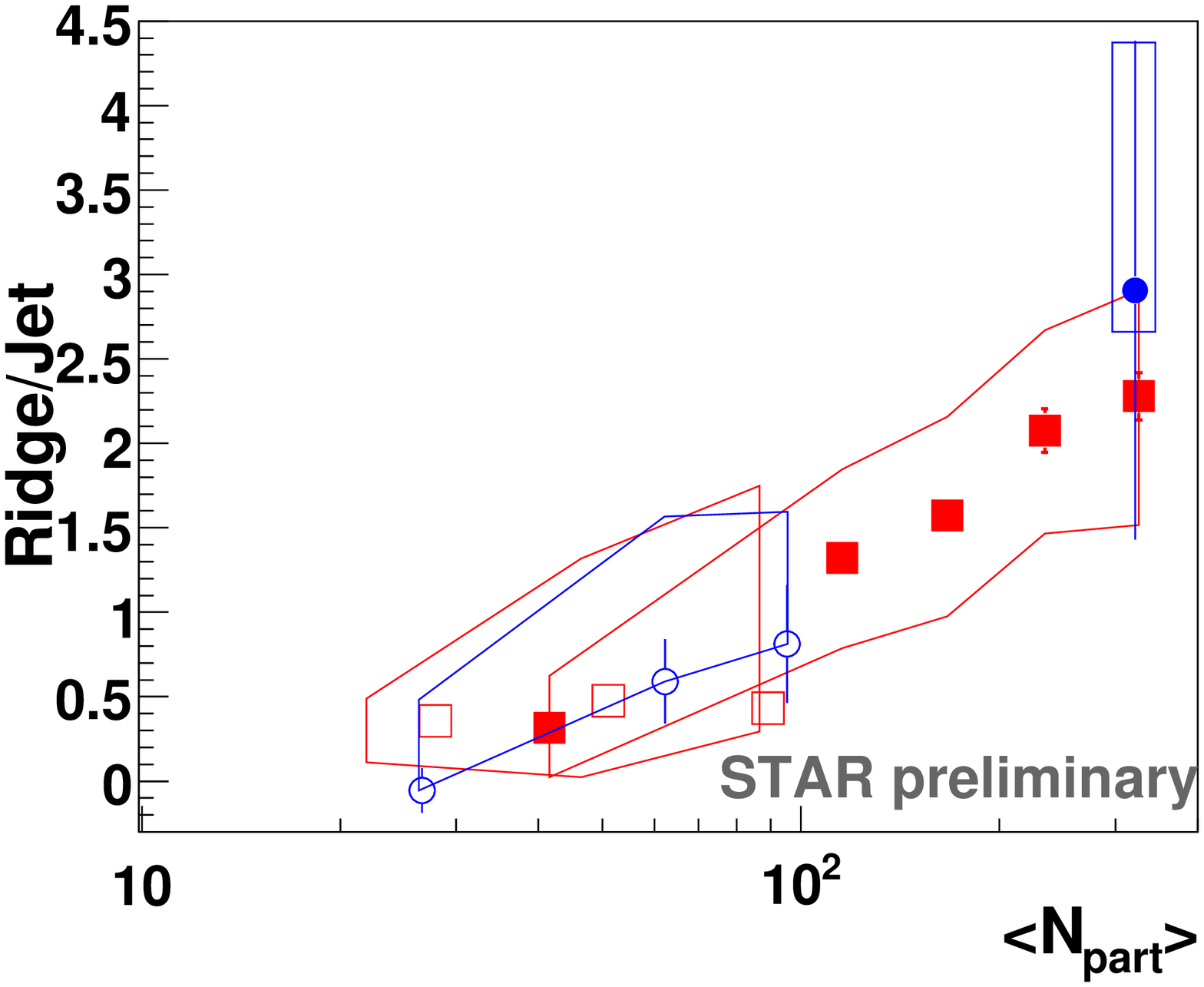}
}}

\vspace{-.3cm} 
\begin{tabular*}{15.cm}{  l@{\extracolsep{4.8cm}}  l @{\extracolsep{4.5cm}} l @{\extracolsep{\fill}}}
\hspace{-0.4cm} \textbf{(c)} & \textbf{(d)} & \textbf{(e)}\\
\end{tabular*}

\vspace{-.3cm} 
\caption{Energy and system dependence of yields for h-h correlations.  Dependence of \jet yield on (a) \pttrig (b) \ptassoc (c) \npart, (d) \ridge yield on \npart (e) \ridge/\jet on \npart.  \stdtrig except for (a) and \stdassoc except for (b).  Data from the fits in (b) are shown in \tref{Table}.  Colour online.}\label{Figure1}
\vspace{-0.9cm} 

\end{center}
\end{figure}

The \jet yield per trigger is shown as a function of \pttrig in \fref{Figure1}a, as a function of \ptassoc in \fref{Figure1}b, and as a function of \npart in \fref{Figure1}c.  The \jet yield at \sNNsixtytwo is considerably lower than that at \sNNtwohundred.  However, similar features are observed at both energies; the \jet yield rises with \pttrig, decreases steeply with \ptassoc, and is independent of \npart.  The data in \tref{Table} demonstrate that the \Cu and \Au inverse slope parameters are within error at the same \sNN and \npart.  The \ridge yield as a function of \npart is shown in \fref{Figure1}d.  The \ridge yield increases dramatically with \npart for both energies.  Although the magnitude of the yield of both the \jet and the \ridge are considerably smaller at \sNNsixtytwo, \fref{Figure1}e shows that \ridge/\jet is roughly independent of energy.

\begin{figure}
\begin{center}
\hspace{0.0cm} 
\rotatebox{0}{\resizebox{15.cm}{!}{
	\includegraphics{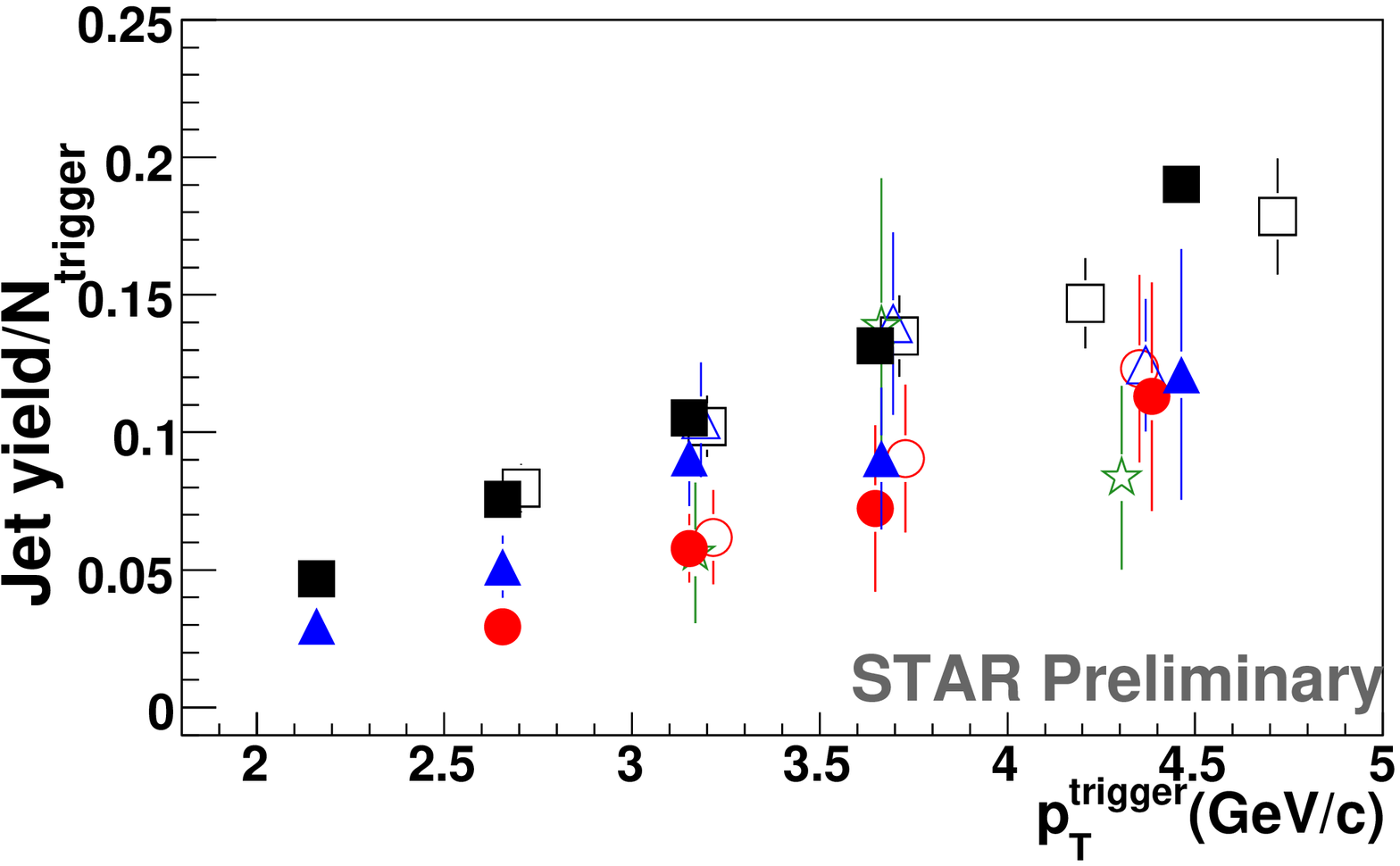}
	\includegraphics{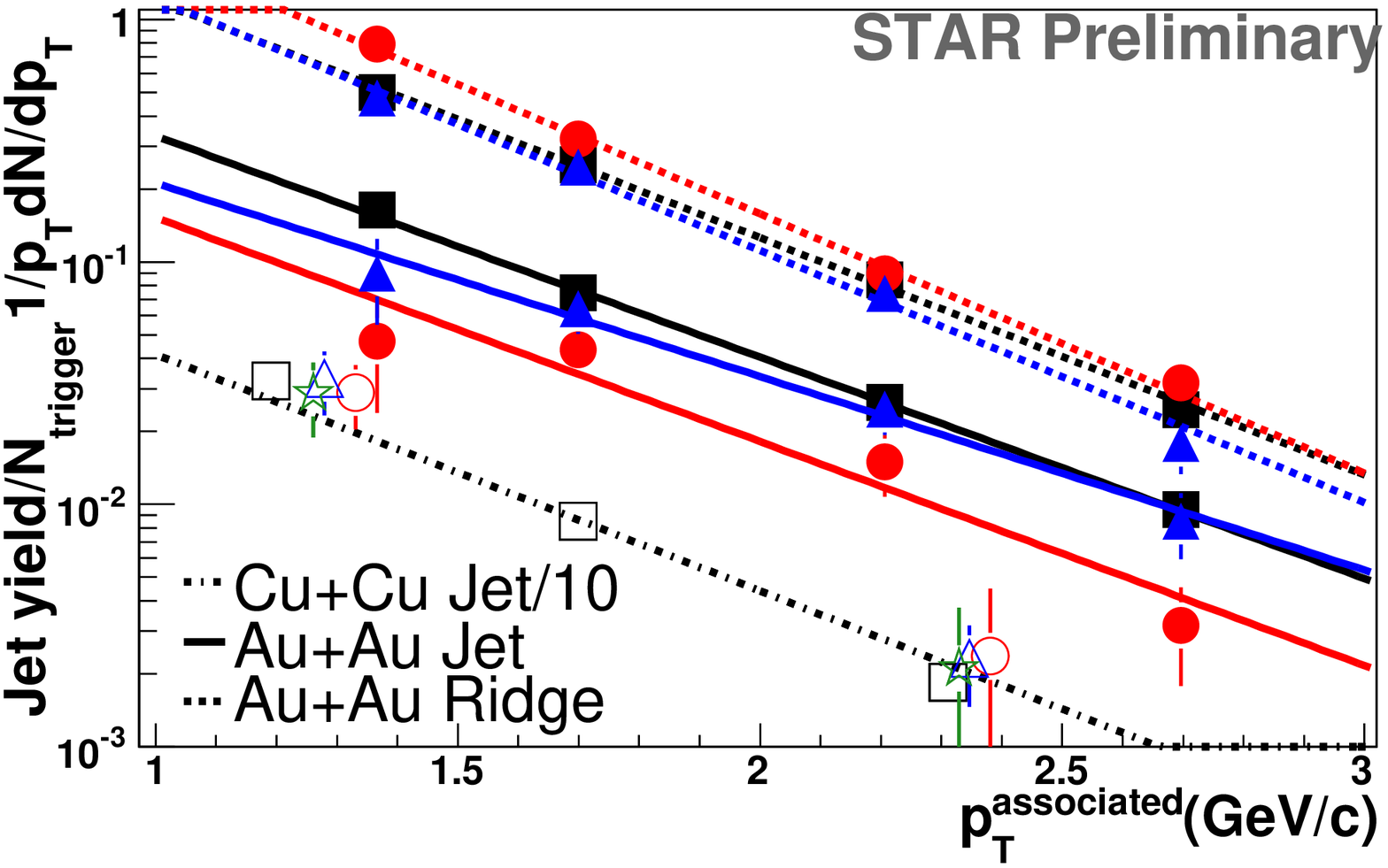}
}}

\vspace{-.6cm} 
\begin{tabular*}{15.7cm}{ @{\extracolsep{6.5cm}} l l @{\extracolsep{\fill}}}
\hspace{0.5cm} \textbf{(a)} & \textbf{(b)}\\
\end{tabular*} 
\rotatebox{0}{\resizebox{15.cm}{!}{
	\includegraphics{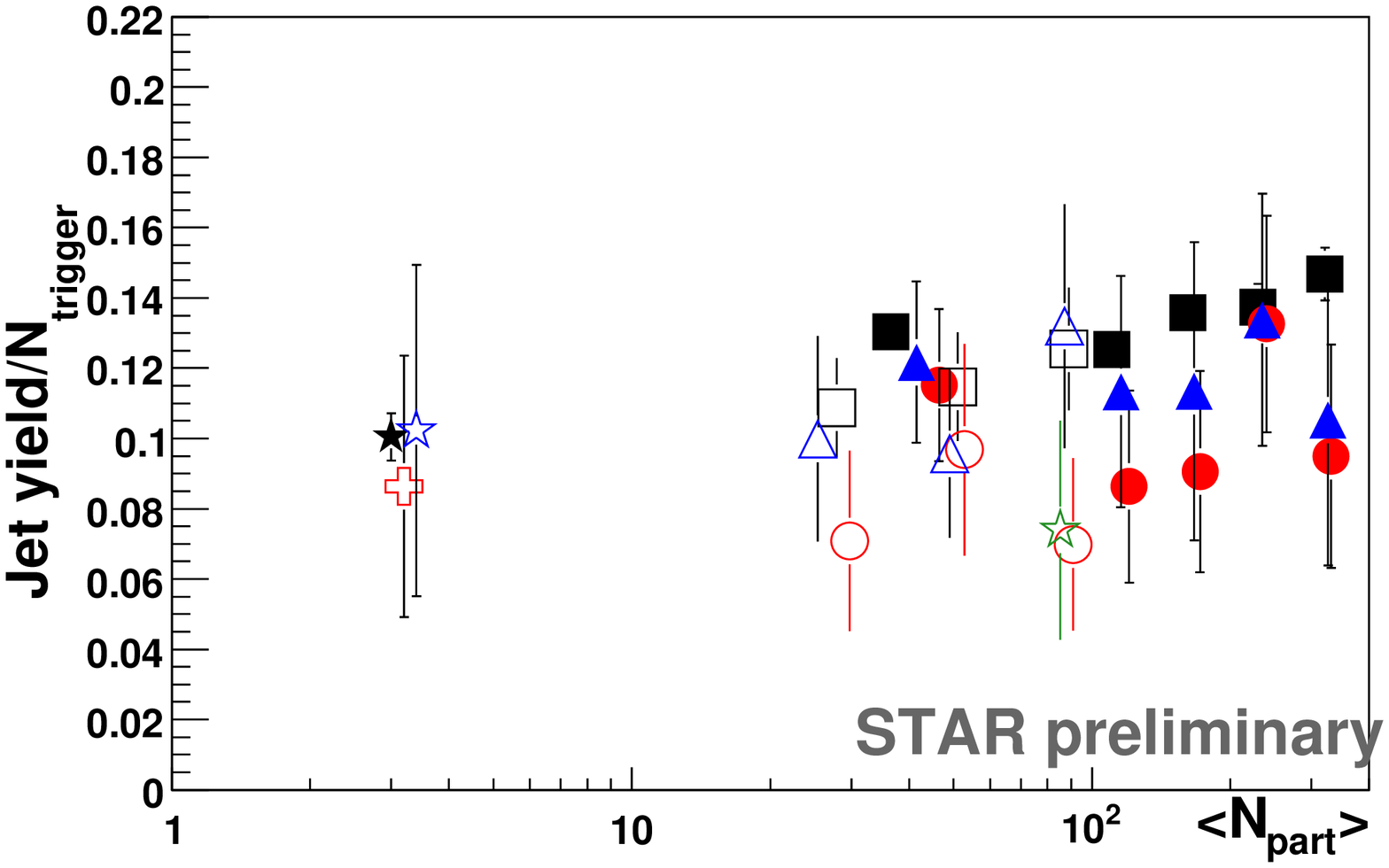}
	\includegraphics{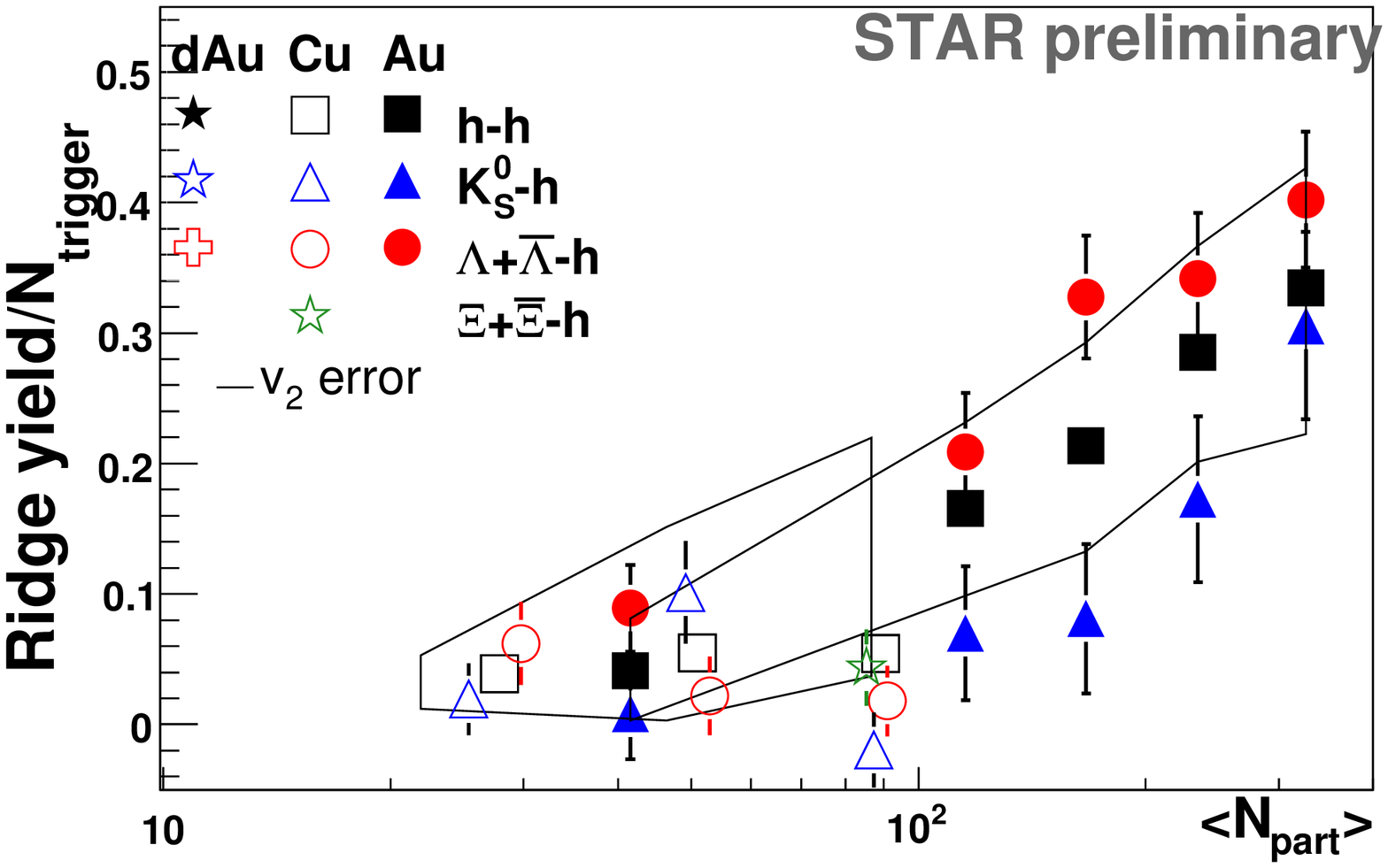}
}}

\vspace{-0.2cm} 
\begin{tabular*}{15.cm}{ @{\extracolsep{6.5cm}} l l @{\extracolsep{\fill}}}
\hspace{0.5cm} \textbf{(c)} & \textbf{(d)}\\
\end{tabular*} 

\vspace{-0.2cm} 
\caption{System and particle type dependence at \sNNtwohundred for identified trigger particles.  (a) Dependence of \jet on \pttrig for \stdassoc (b)  Dependence of \jet and \ridge on \ptassoc for \stdtrig (c) Dependence of \jet on \npart (d) Dependence of \ridge on \npart.  Data from the fits in (b) are shown in \tref{Table}.  The systematic errors due to \vtwo are comparable to those shown for unidentified triggers for \kaon triggers and roughly 3/2 times larger for \lam and \xibary\cite{Jana,Me}.  Colour online.}\label{Figure2}
\vspace{-0.9cm} 
\end{center}
\end{figure}

\Fref{Figure2} shows the yields with identified trigger particles at \sNNtwohundred.  The dependence of the \jet yield on \pttrig, \ptassoc, and \npart, shown in \fref{Figure2}a, b, and c, respectively, may indicate a slightly lower yield for \lam and \kaon.  However, the systematic error due to tracking inefficiencies for close pairs, which is greater for \lam and \kaon triggers than for h, may be sufficient to explain the observed difference.  \Fref{Figure2}b and \fref{Figure2}d show the dependence of the \ridge on \ptassoc and \npart, respectively; these data may indicate a slightly higher yield for \lam and a slighly lower yield for \kaon triggers.  However, the data are within error for all trigger species and the degree of correlation of errors for different trigger particle species is still under study.  Therefore, within our current understanding of the errors, there is no significant dependence on the trigger particle species in either \Cu or \Au collisions.  The independence of the inverse slope parameters in \tref{Table} on particle species further supports no trigger type dependence.

\begin{figure}
\begin{center}
\hspace{0.0cm} 
\rotatebox{0}{\resizebox{15.cm}{!}{
	\includegraphics{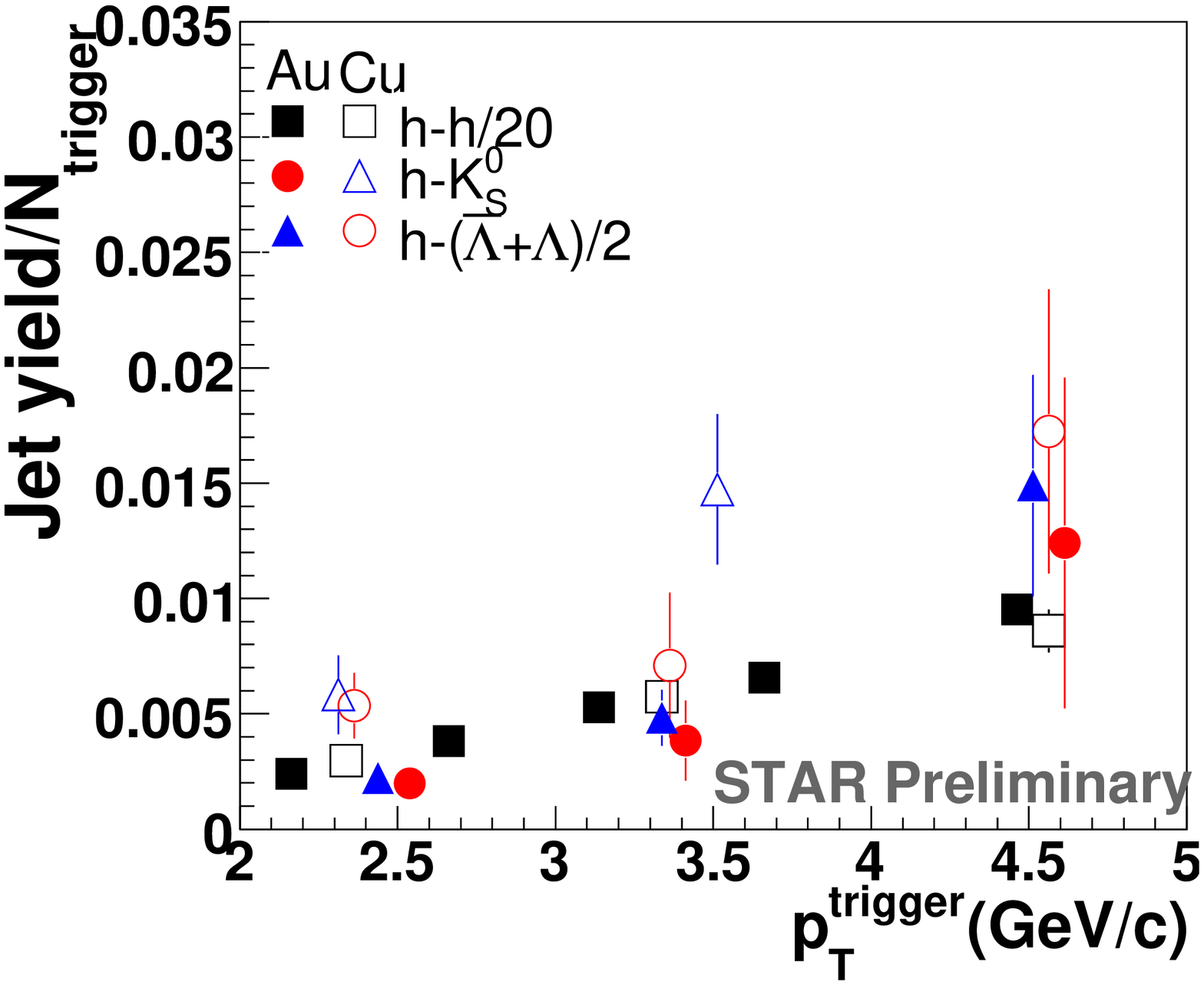}
	\includegraphics{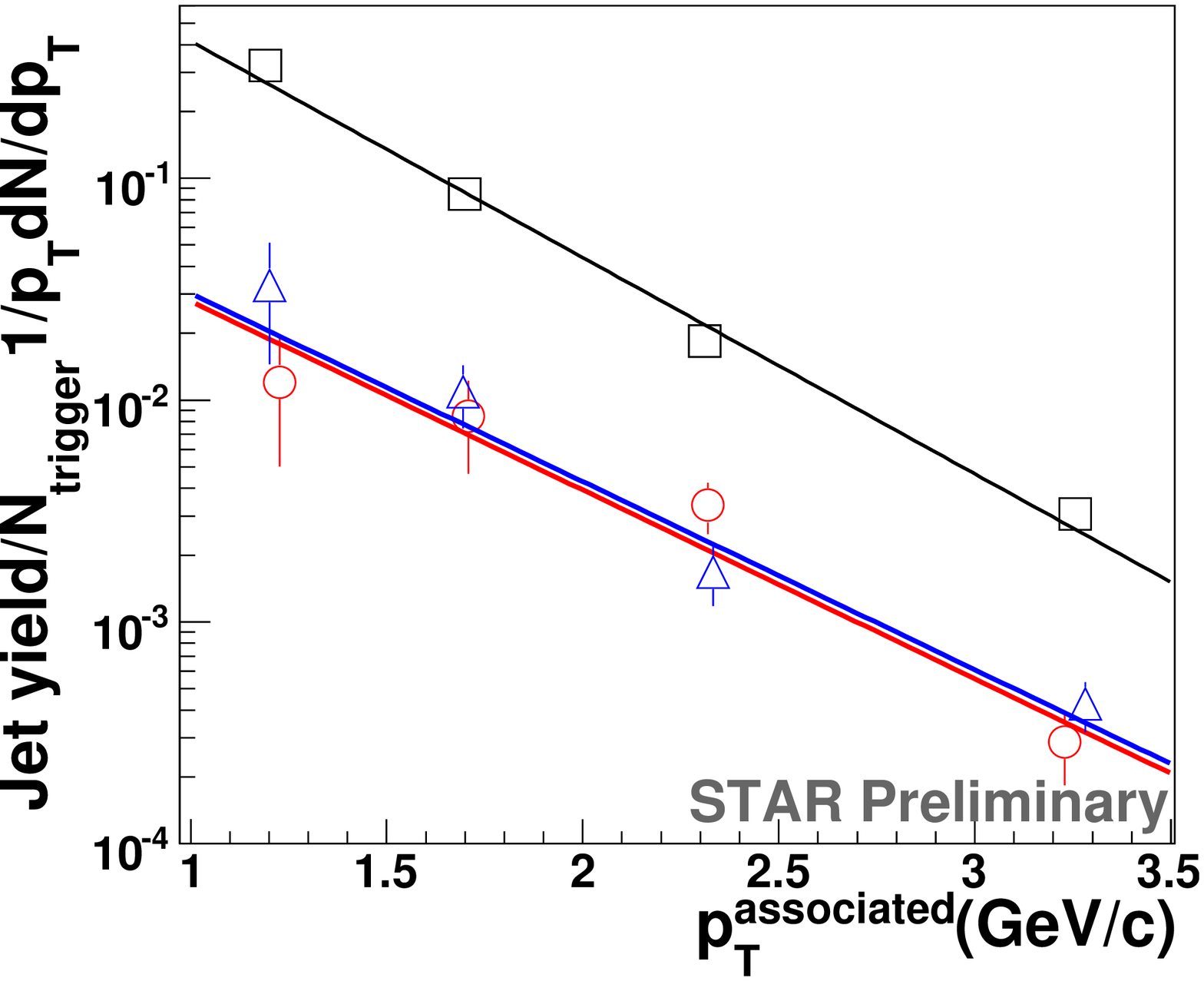}
	\includegraphics{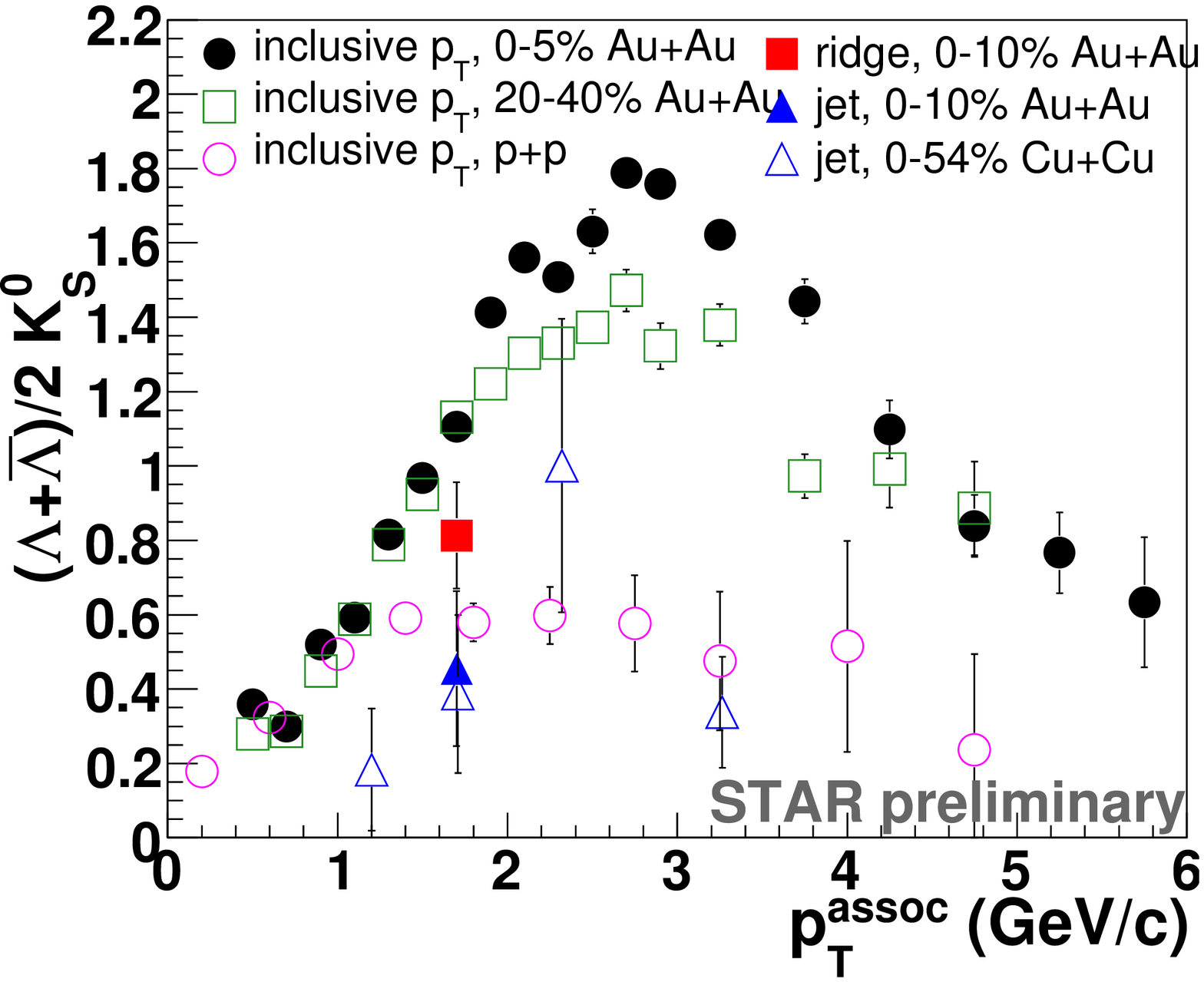}
}}

\vspace{-.5cm} 
\begin{tabular*}{15.cm}{  l@{\extracolsep{4.8cm}}  l @{\extracolsep{4.5cm}} l @{\extracolsep{\fill}}}
\hspace{-0.4cm} \textbf{(a)} & \textbf{(b)} & \textbf{(c)}\\
\end{tabular*} 

\vspace{-.3cm} 
\caption{System and particle type dependence of the \jet at \sNNtwohundred for identified associated particles.  (a)  Dependence on \pttrig for \stdassoc (b) Dependence of \ptassoc for \stdtrig  (c) Particle ratios in the \ridge and \jet as compared to the inclusive particle ratios.   Data from the fits in (b) are shown in \tref{Table}.  Colour online.}\label{Figure3}
\vspace{-0.9cm} 
\end{center}
\end{figure}

\Fref{Figure3} shows the yields with identified associated particles at \sNNtwohundred.  The dependence on \pttrig shown in \fref{Figure3}a and on \ptassoc in \fref{Figure3}b for \Cu and \Au collisions at \sNNtwohundred are similar for both \kaon and \lamalam to that observed for unidentified hadrons, although the magnitude is considerably smaller.  The baryon to meson ratios in the \ridge in \Au, the \jet in \Au and \Cu, and the inclusive spectra in \Au and \pp are compared in \fref{Figure3}c.  The baryon to meson ratio in the \ridge is within error of that of the inclusive ratio in \Au while the baryon to meson ratios in the \jet in \Au and \Cu are comparable to that in \pp.

\begin{table}
\begin{center}
\caption{Inverse slope parameter k (\MeV) of \ptassoc for fits of data in \Fref{Figure1}(b), \Fref{Figure2}(b), and \Fref{Figure3}(b) to $A e^{-p_T/k}$.  The inverse slope parameter from a fit to $\pi^-$ in \Au from \cite{Pion} above 1.0 \GeV is $k$ = 280.9 $\pm$ 0.4 \MeV for \sNNsixtytwo and is $k$ = 330.9 $\pm$ 0.3 \MeV for \sNNtwohundred.  Statistical errors only.}\label{Table}
\hspace{-.2cm}

\begin{tabular}[b]{c|c|ccccc}
\hline
 & 62 $GeV$ & \multicolumn{5}{c}{200 $GeV$}\\
   & h-h & h-h & \kaon-h & \lamalam-h & h-\kaon & h-\lamalam\\ \hline
Au \ridge&  & 438 $\pm$ 4 & 406 $\pm$ 20 & 416 $\pm$ 11 & &  \\ 
Au \jet & 317 $\pm$ 26 & 478 $\pm$ 8 & 530 $\pm$ 61 & 445 $\pm$ 49  & &\\
Cu \jet & 355 $\pm$ 21 & 445 $\pm$ 20 & & & 505 $\pm$ 87 & 510 $\pm$ 55 \\ \hline
\end{tabular}
\vspace{-0.8cm} 
\end{center}
\end{table}

\section{Conclusions}
The measured \jet and \ridge yields in \Cu and \Au collisions are within error at the same \sNN and \npart, indicating no dependence on geometry.  The \jet yield at \sNNsixtytwo is smaller than at \sNNtwohundred, as expected from the steeper spectrum in \sNNsixtytwo\cite{Kirill}; the observation that the \ridge/\jet ratio is roughly independent of energy is potentially a powerful way to distinguish models.  No dependence on the trigger particle identity is observed.  Baryon/meson ratios in \jet and \ridge and the differences in the associated particle spectra suggest that the \jet has properties similar to \pp, whereas the properties of the \ridge are closer to the bulk.


\end{document}